# MeV-peaked electrons in a coronal source of a solar flare

Gregory D. Fleishman[1,2*], Ivan Oparin[1], Gelu M. Nita[1], Bin Chen[1], Sijie Yu[1], Dale E. Gary[1]

[1]Center for Solar-Terrestrial Research, New Jersey Institute of Technology, Newark, NJ 07102, USA.
[2]Institut für Sonnenphysik (KIS), Georges-Köhler-Allee 401 A, D-79110 Freiburg, Germany.

*Corresponding author(s). E-mail(s): gfleishm@njit.edu;

**Abstract**

Analysis of $\gamma$-rays in solar flares has suggested a distinct continuum component dominating at MeV energies, which differs from the well-studied X-ray continuum produced by flare-accelerated electrons with spectra steeply falling with energy. The origin, precise spatial location, and extent of this mysterious MeV component have been unknown up to now. If it is produced by bremsstrahlung, such a $\gamma$-ray component requires an unusual population of electrons peaked at a few MeV. Here we report a joint study of this MeV-peaked electron population in the 2017-Sep-10 solar flare with *Fermi* MeV $\gamma$-ray data and EOVSA spatially resolved microwave imaging spectroscopy data. We demonstrate that the microwave spectrum from the MeV-peaked distribution has a distinctly different shape from that produced by the electrons with falling energy spectrum. We inspected microwave maps of the flare and identified an evolving area where the measured microwave spectra matched the theoretically expected one for the MeV-peaked population, thus pinpointing the site where this MeV component resides. The locations are in a coronal volume adjacent to the region where prominent release of magnetic energy and bulk electron acceleration were detected, which implies that transport effects play a key role in forming this population.

Solar flares promptly release large amounts of free magnetic energy in the solar corona [1, 2] to produce substantial populations [3, 4] of high-energy charged particles, both ions and electrons. These particles are detected when they radiate microwaves in solar magnetic fields and X- and $\gamma$-rays when they encounter matter. Over decades it was customary to study high-energy electrons energized in solar flares using the bremsstrahlung they produce in X-ray and $\gamma$-ray domains due to collisions with ambient particles. This approach is very robust because the spectral shape of the bremsstrahlung is uniquely linked with the spectral shape of the electron energy spectrum (e.g. the slope in the case of power-law spectra). Typically, the X-ray and $\gamma$-ray spectrum falls with increasing photon energy, first steeply up to several hundred keV, flattening up to several MeV, and then falling steeply again [5–9]. The key question is what is the energy spectrum of electrons that gives rise to this photon spectrum.

The energy distribution of electrons at energies up to a few hundred keV in solar flares is well determined by hard X-ray observations. At higher energies, their energy distribution is more uncertain, especially from about 500 keV to 10 MeV. This uncertainty is primarily due to the addition of nuclear $\gamma$ radiation in the energy range produced by flare-accelerated nuclei that interact in the solar chromosphere. However, with current knowledge of the cross sections for producing nuclear $\gamma$-rays in the solar atmosphere, their contribution to the solar spectrum is now sufficiently well determined that reliable subtraction is possible [9]. After subtraction, the revealed residual continuum often displays an unexpected flattening of the photon spectrum above several hundred keV [5–9]. Possibilities to account for this spectral flattening are (i) a corresponding flattening of the electron energy spectrum [8], (ii) a separate, energy-peaked population of electrons [9], or (iii) a different gamma-ray emission process than bremsstrahlung, e.g., inverse Compton scattering [7, 9]. A recent comprehensive study [9] considers option (i) and argues that hardening of a single source of electrons is inconsistent with observed differences in timing, directionality, and location. Conclusive evidence to exclude this possibility is not possible, though, based on the $\gamma$-ray data alone. That study also could not unambiguously select between options (ii) and (iii). To account for the photon spectrum via the inverse Compton mechanism requires a flat electron energy spectrum that rolls over between 10 − 20 MeV, while for bremsstrahlung the energy spectrum of this new electron population must rise to a peak between 2 and 6 MeV and then fall rapidly at higher energies [9]. For the 2017-Sep-10 flare considered here, the $\gamma$-ray inferred electron spectrum [9] rises as $\propto E^{\delta_1}$ with $\delta_1 \leq 2$ up to the break energy $E_{\rm break} = 1 - 5\,{\rm MeV}$ and then rapidly falls as $\propto E^{-\delta_2}$ with $\delta_2 = 2 - 8$ (see Methods and fig. 9 from [9]).

Such an electron population would be a strong contributor to the microwave emission due to gyro-magnetic processes in the ambient magnetic field, and so should be directly detectable in spatially resolved maps of the microwave brightness. Indeed, the brightness temperature[1] (in energy units) of this microwave emission could be up to a few MeV in the optically thick regime; thus, observing microwave emission with high brightness temperature is a direct signature of charged particles with the same or larger energy.

[1] Microwave brightness temperature is a measure of the intensity of microwave radiation emitted by an object, expressed in terms of the equivalent temperature of a blackbody that would emit the same amount of radiation at a given wavelength or frequency.

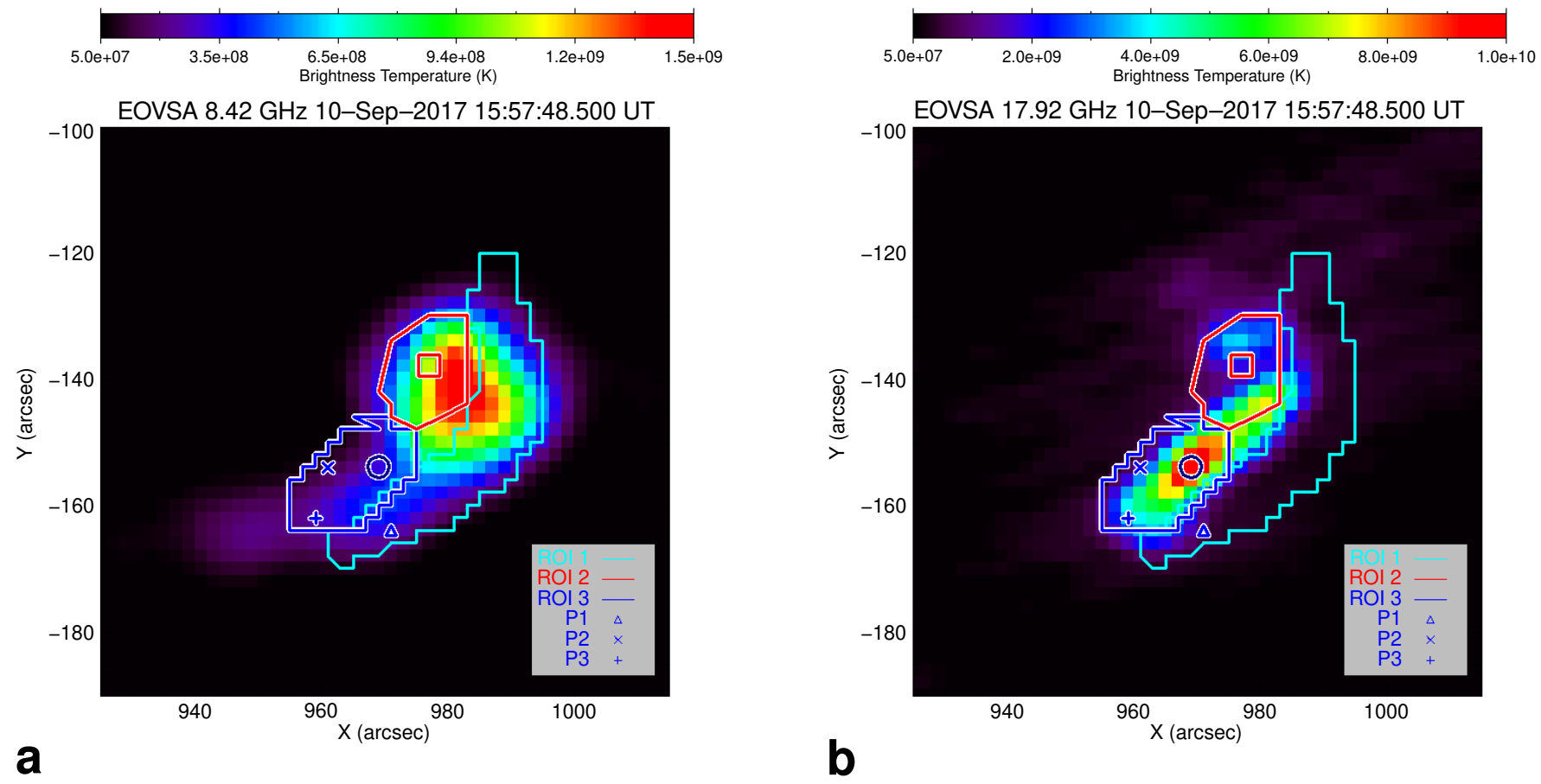

**Fig. 1 Microwave maps of brightness temperature.** The distribution of the brightness temperature of microwave emission observed with EOVSA at (a) 8.42 and (b) 17.92 GHz at 15:57:48 UT along with three different ROIs defined in the legend (cyan, red, and blue lines for ROI 1, 2, and 3, respectively). The images show different morphologies. The brightness temperature of the high-frequency source (b) has a peak value of $10^{10}$ K indicative that MeV particles are the main contributors to this emission. The red square and blue circle symbols mark the selected pixels corresponding to the spectra presented in Figure 2; triangle, cross, and plus symbols correspond the P1, P2 and P3 pixels for which the MCMC results are shown in Extended Data Figures 3, 5, and 4, respectively.

Figure 1 shows images of microwave emission observed with EOVSA at two frequencies, 8.42 and 17.92 GHz, at 15:57:48 UT as an example. These images are morphologically different, with lower-frequency emission peaking in ROI 1 and ROI 2 [10], while higher-frequency emission peaks in ROI 3. The brightness temperature $T_B$ in the high-frequency source (right panel) is very large, with the peak value of $T_B \approx 10^{10}$ K. Such a high brightness temperature for incoherent continuum microwave emission at these high frequencies is direct evidence that relativistic MeV particles are the main contributors of this emission. Indeed, if this emission is optically thick, the brightness temperature is equal to the mean energy of the particles that form the gyrosynchrotron (GS) opacity, which is about 0.9 MeV for $T_B = 10^{10}$ K. This also implies that the opacity due to lower-energy electrons is negligible; thus, these lower-energy particles are not numerous in this region. However, as our spectral analysis shows below, the emission is optically thin, which requires even higher-energy particles to produce the observed microwave emission.

Figure 2 shows two contrasting examples of spatially resolved microwave spectra—one from the center of ROI 2 introduced in [10] (red symbols) and one from a pixel in ROI 3 (blue symbols). A zoomed-in version (panel b) of these spectra clearly shows a large difference in the spectral slopes. Indeed, the red spectrum rises as $\propto f^{\beta}$, with an index $\beta \approx 2.7$, typical for the optically thick GS emission from a power-law distribution of nonthermal electrons accelerated in flares. The red curve shows a model spectral fit to the (red) data points assuming a conventional power-law distribution of nonthermal electrons with spectral index $\delta = 2.6$ between $E_{\rm min} = 20$ keV and $E_{\rm max} = 5$ MeV.

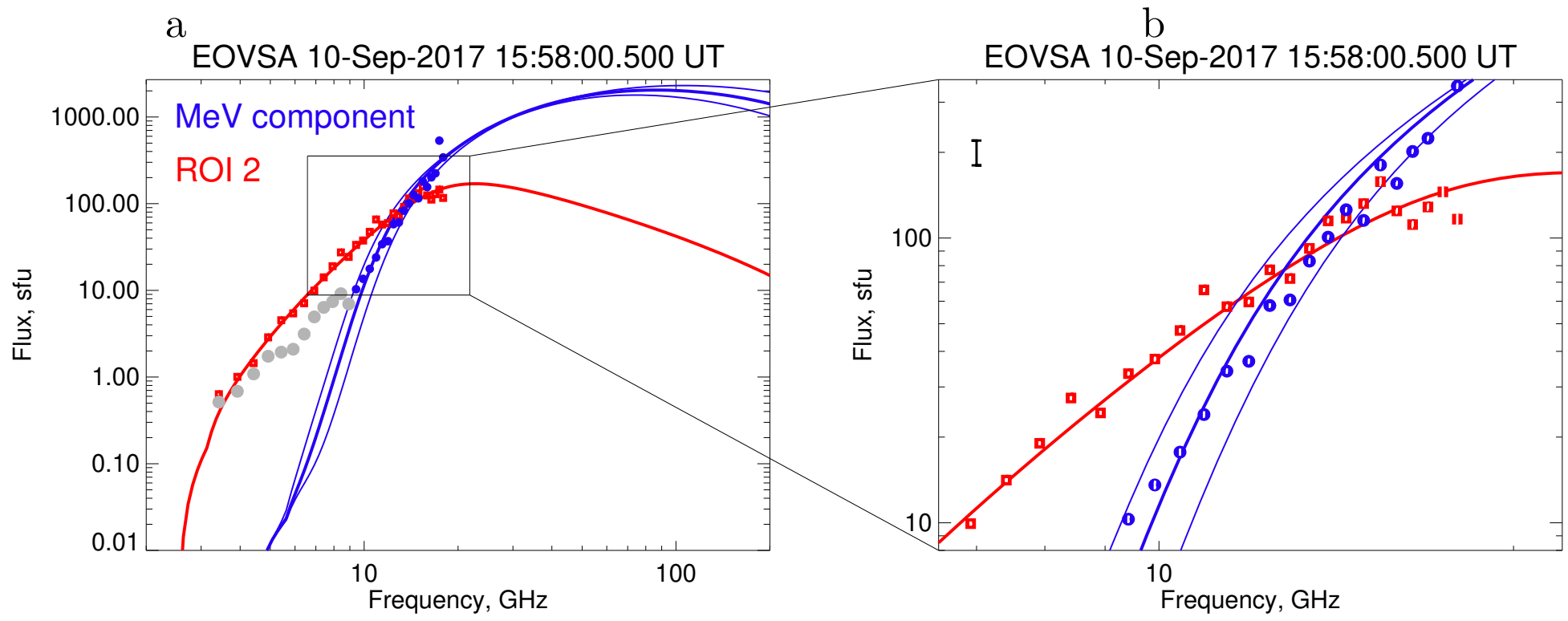

**Fig. 2 Example of "normal" and distinctly different (abnormally steep) microwave spectra.** (a) Red squares show a normal optically thick GS spectrum from a central pixel within ROI 2 [10] blue circles show an example of an abnormally steep high-frequency spectrum from ROI 3 introduced in this paper, while light gray circles show lower-frequency data from the same pixel with a shallower spectrum, presumably unrelated to the high-frequency steep component. The statistical uncertainties (not shown in this panel) are about or less than the symbol sizes. (b) A zoomed-in version of the same plot emphasizing the distinctly different slopes of the spectra. The statistical uncertainties are shown as white vertical bars on top of the symbols, while an estimate of systematic uncertainty due to calibration error is shown by a black vertical error bar. The red lines in both panels show the spectral model fit to the red data points using a power-law distribution of nonthermal electrons extending from 20 keV to 5 MeV with $\delta = 2.6$ ($B = 420$ G, $n_{th} = 7.6 \times 10^{10}$ cm$^{-3}$). The set of blue lines shows the GS spectra from the MeV-peaked component with a power-law electron distribution with $\delta = 2$, for three different values of $E_{\rm min}$ (2, 2.6, 3.6 MeV) extending to $E_{\rm max} = 6$ MeV ($B = 750$ G, $n_{th} = 2.46 \times 10^{11}$ cm$^{-3}$).

In contrast, the blue symbols follow a much steeper slope between 8 and 18 GHz with the spectral index $\beta \approx 6$. Such a steep slope is inconsistent with the normal GS optically thick spectrum for which $\beta \lesssim 3$ must hold [11]. This steep spectrum can be formed when GS opacity goes down quickly towards lower frequencies due to lack of sub-MeV and keV electrons, which would otherwise provide a substantial opacity, while the emissivity and opacity of relativistic MeV particles is suppressed by dense ambient plasma due to Razin-effect [12]; see also supplementary video in [10] and the Methods Section below. Accordingly, we calculate the blue curves showing GS spectra from a power-law MeV component with three different values of low-energy cutoff $E_{\rm min} = 2, 2.6, 3.6$ MeV, quantitatively consistent with the electron spectrum inferred from the $\gamma$-ray data (see Methods). These model spectra are a good match to the observed one, which is indicative that the same MeV-peaked electrons (or positrons) are responsible for both the flat gamma-ray continuum component and the abnormally steep-spectrum GS emission. The MCMC model spectral fitting (see Methods) confirms that an MeV-peaked component is indeed needed to fit these abnormally steep spectra.

Capitalizing on this finding, we employ this abnormally steep, exceptionally bright high-frequency microwave emission to locate the mysterious MeV electron/positron component in space and investigate its properties and evolution. Figure 3a displays multifrequency microwave brightness above $T_B = 300$ MK at multiple times, which shows the appearance and disappearance of this component (blue areas), temporally well correlated with the MeV-peaked electron component shown in panel b, derived from the $\gamma$-ray spectral fit reported in the Methods Section.

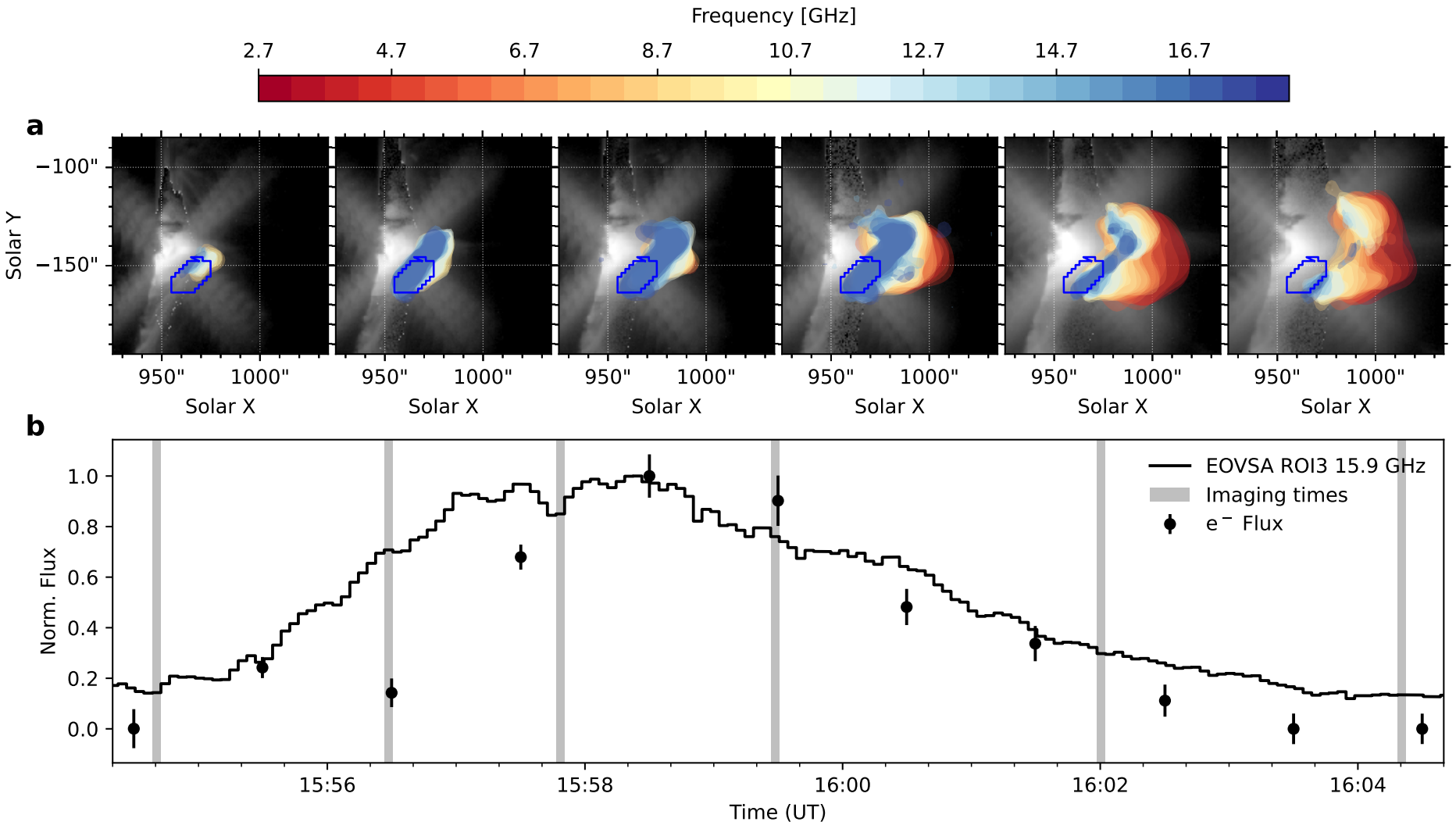

**Fig. 3 The relationship between microwave emission and electron flux.** (a). Multi-frequency EOVSA and AIA 131 Å image sequences illustrating the spatial and spectral evolution of microwave emission at times marked by the vertical bars in (b). Overlaid filled contours shown brightness temperature over 300 MK at multiple frequencies (colors denote frequency; see legend). The blue contour outlines ROI 3. (b) Comparison of the time history of the normalized MeV electron flux (with error bars showing ±n SD or ±n SEM—Gregory, please confirm how these were calculated) and the 15.9 GHz flux from ROI 3 during the flare.

We identified pixels in the spatially resolved microwave images whose peak brightness temperature exceeds $10^9$ K, and performed a spectral fitting of such pixels for frequencies above 8 GHz, using a simple rising power-law function $F \propto f^\beta$ with index $\beta$. Figure 4 shows a map of this spectral index for a single time (an animation covering the whole analyzed flare duration is available), where the diverging color table cleanly separates cases with $\beta < 3$ and $\beta > 3$. It is apparent that the abnormally steep spectra with $\beta > 3$ cluster in/around ROI 3. The animation shows that this green-color region appears at the left part of ROI 3, then moves to the right (i.e., away from the solar surface) with time, as the entire system expands, and then fades away.

The presented evidence allows us to pinpoint the spatial location, ROI 3, and the evolving extent of this MeV-peaked component in the morphological context of the flare. The ROI 3 volume includes a mixture of the MeV-peaked component and the usual dense flaring thermal plasma. It is spatially distinct, but adjacent to two other key regions of this flare [10], ROI 1 and 2. ROI 1 is the region where a fast decay of the magnetic field has been detected [2] accompanied by highly efficient acceleration of literally all thermal electrons to high energies [10]. ROI 2 is a region filled with a more usual dense thermal plasma with a modest population of nonthermal electrons with falling spectrum. Therefore, these three ROIs display highly contrasting properties

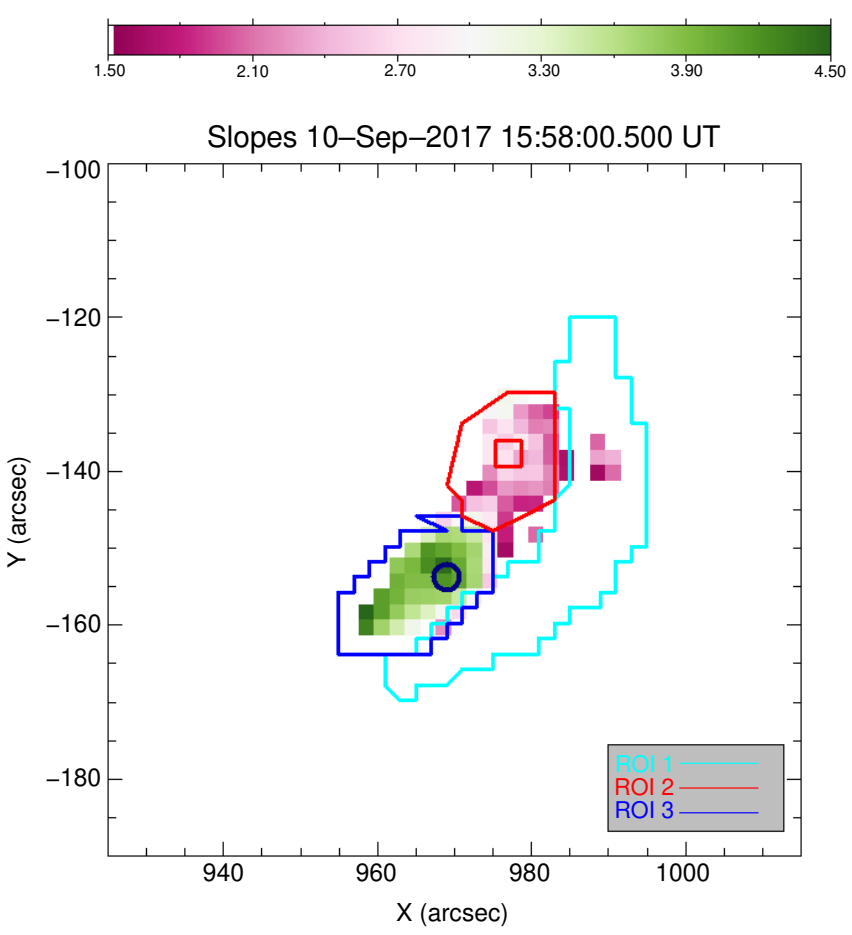

**Fig. 4 Map of the microwave emission spectral indices.** The map is masked to the pixels with brightness temperature above $10^9$ K. Color contours show three ROIs (1, 2, 3) same as in Fig. 1. Two symbols mark the pixels whose spectra are shown in Fig. 2. It is apparent that the abnormally steep spectra (indicated by the green color) are concentrated in ROI 3. Animation is available.

of the nonthermal populations and strongly divergent thermal-to-nonthermal particle partitions.

We use this information to discuss the possible origin of this component. We envision several scenarios of this origin: (i) the charged particles, electrons and/or positrons, are already created with (sub-)MeV energies due to $\beta^{\pm}$ decay of radioactive nuclei or decay of $\pi^{\pm}$ pions created by flare-accelerated ions; (ii) a peculiar acceleration mode/mechanism, which somehow creates mainly MeV electrons, but not keV ones; or (iii) evolution of flare-accelerated electrons with falling spectrum, where low-energy electrons are quickly lost, while MeV electrons survive longer.

Option (i) is attractive since high-energy positrons/electrons are already created in the $\beta^{\pm}, \pi^{\pm}$ decay processes and, thus, no additional acceleration process might be needed. Yet, we do not find compelling evidence for this model as most of the $\beta$-decay nuclei have a rather long decay time [13–15], which seems inconsistent with the timing of the $\gamma$-ray emissions produced by the MeV component. In contrast, the prompt high-energy, $> 100\,\mathrm{MeV}$, $\gamma$-ray emission from $\pi^0$-decay was observed in this flare to be overall co-temporal with the MeV-peaked component [16]. Moreover, this component displayed the largest $\pi^0$-decay emission flux detected over the history of Fermi/LAT observations [17]. However, the pion production requires interactions in a dense (sub-)photospheric layer, presumably much denser than the coronal source located at ROI 3. For this option to work requires an efficient escape of the secondary positrons from the photosphere up to the corona, for which we do not have any evidence. A crucial test to distinguish microwave emission from relativistic electrons and positrons would be to observe circular polarization of the microwave emission, which has opposite sense for positive and negative emitting particles [18]. No polarization data are available for this event.

Option (ii) requires rather special conditions, where only a minor fraction of the thermal electrons, $\sim 10^{-6}$ gets accelerated but to very high, MeV, energies. This process is unlikely to be a stochastic mechanism, as such a process would produce energy spectra that fall with energy; thus, the energy gain must be direct in our case. To accelerate a $10^{-6}$ portion of the thermal particles (see Methods), we need an electric field $E \approx E_D/28$, where $E_D$ is the Dreicer field [19], which is $E_D \approx 1.5 \times 10^{-3}\,\mathrm{V\,cm^{-1}}$ for the parameters obtained for ROI 3. To transfer the energy of 2 MeV to an electron, this electric field has to do work on it equal to $W = \int E ds \approx \langle E \rangle S \approx 2\,\mathrm{MeV}$, where $S$ is the electron path, which requires $S \approx 4 \times 10^{10}\,\mathrm{cm}$ for $E \approx E_D/28 \approx 5 \times 10^{-5}\,\mathrm{V\,cm^{-1}}$. This length is more than a factor of 10 larger than ROI 3 size and, thus, can only be achieved due to strong spatial diffusion, which implies a stochastic trajectory, which is in contradiction with the requirement of direct acceleration; thus, ruling out this option; see Methods.

Option (iii) implies a modification of the electron spectrum with time that leads to an MeV-peaked distribution. Why might the lower-energy electrons be missing from ROI 3? Possibly, there was initially a standard falling electron population there because of the release of the free magnetic energy and associated particle acceleration. Once this free energy is expended, the electrons would experience a collisional evolution similar to that described in [20] resulting in the loss of low-energy electrons, which is plausible given that the coronal density up to $\sim 10^{12}\,\mathrm{cm^{-3}}$ has been reported for this flare [21].

Evolution of the charged particle distribution is controlled by a balance between energy gain and loss. Collisional evolution of a trapped power-law electron population results in an energy spectrum with a spectral peak rising with time [22]. This happens because the collisional loss time rises with the electron energy as $\propto E^{3/2}$ in the nonrelativistic domain and, thus, low-energy electrons lose their energy much faster than the high-energy ones. For the estimated thermal number density of $\sim 2.5 \times 10^{11}\,\mathrm{cm^{-3}}$, see caption to Fig. 2 and the Methods section, the Coulomb loss time for 1 MeV electrons is about 30 s. This implies that if ROI 3 was initially filled with a standard power-law electron distribution with a falling spectrum, then, in the absence of additional acceleration, the sub-MeV electrons will be lost in about 30 s and an MeV-peaked distribution can be formed. The Coulomb losses dominate for electrons up to 7-9 MeV, while then synchrotron losses become dominant for the ROI 3 magnetic field estimated at 400-500 G. The electrons with an energy of several MeV, without additional acceleration or injection, will survive at a given location for about two minutes, which is comparable to the total duration of this MeV-peaked component, $\sim 8$ minutes. They are consistent with each other given the evolution (upward motion) of the source.

If this MeV-peaked electron component forms due to this collisional evolution of the coronal trapped population of the nonthermal electrons, as we argued above, the formation of such a component must be common and not limited to only the strongest flares. This is consistent with the finding [9] of a similar flat $\gamma$-ray component in a summed spectrum of 48 flares, which individually did not show it because of insufficient statistics. To conclude, our study removes the ambiguity of the emission mechanism responsible for the $\gamma$-ray continuum by confirming the presence of the MeV-peaked electron component needed for the bremsstrahlung mechanism to work in the coronal flare source and also identifies the source region and extent of this new electron component.

# Methods

## Overview

We employ joint observations of the 2017 September 10 solar flare by the *Fermi* Gamma-ray Burst Monitor (GBM) and the EOVSA to compare the microwave and $\gamma$-ray emission from an MeV-peaked component of electrons. EOVSA has studied spatially and spectrally resolved emission from several regions of this flare [2, 10] distinct from the ROI 3 region identified here. The *Fermi*/GBM observed $\gamma$-ray emission from this flare lacks spatial resolution but provides energy spectra and time evolution.

## Spectral fitting of *Fermi*/GBM $\gamma$-ray data

As a starting point, we consider the spectral fit results [9], who reported the three key components of the solar $\gamma$-ray emission in the flare spectrum: (i) the power-law extension of the hard X-ray bremsstrahlung in the form $F_{PWL} = (\alpha - 1)F_0(> E_0)E_0^{\alpha-1}/E^{\alpha}$, where $F_0(> E_0)$ is the photon flux above $E_0$, $E_0 = 300\,\mathrm{keV}$; (ii) the nuclear-line radiation in the form of a standard template composed of narrow and broad $\gamma$-ray lines, and (iii) thin-target bremsstrahlung from the MeV-peaked electron component in a form of broken power-law quantified by the total electron flux $F_{MeV}$ [electrons cm$^{-2}$ s$^{-1}$], break energy $E_{\mathrm{break}}$ [MeV], and low- and high-energy spectral indices, $\delta_{1,2}$ and their uncertainties. The uncertainties of the reported spectral indices [9] are rather large; therefore, they are not well-constrained, although the low-energy index $\delta_1$ is confidently negative, which means a spectrum that increases with energy up to $E_{\mathrm{break}}$.

Clearly, the adopted split of the observed $\gamma$-ray spectrum onto these three components is not unique, so other fitting models can be employed. Here, we do not attempt to identify the best possible fitting model, but rather follow the choice adopted in [9], which took into consideration not only the spectral shape but also other distinct properties of these components, for the entire available time range to recover time evolution of the involved model parameters. We extended the spectral fit [9] to the entire time when this MeV component is present using OSPEX spectral analysis software [23] by considering the three spectral components mentioned. For the power-law component, we allowed the normalization constant and the slope $\alpha$ to vary, while for the thin-target model (`thin2`) we fixed the indices and the break energy at the best-fit reported values [9] (as they are anyway determined with rather large uncertainties) and kept the electron flux as the free parameter. The nuclear template contains only one free parameter, the normalization; therefore, we applied a fit with four free parameters to the Fermi data using one-minute time bins.

We used the measurements from *Fermi*/GBM bismuth germanate (BGO) detector covering the energy range from 150 keV to 30 MeV. Extended Data Figure 1 shows 1-min integrated *Fermi* GBM/BGO spectra near the peak of the emissions and 5 minutes later. During that time interval, the power-law fitting component, which represents the extension of bremsstrahlung continuum in hard X-rays, has hardened significantly from an index of 3.2 to an index of 2.2. In addition, both the bremsstrahlung from the MeV electron component and the nuclear-line emission decreased faster relative to the power-law emission by the end of the analyzed episode. While the `thin2` function component in OSPEX software does not currently include the electron-electron bremsstrahlung, we checked, using a direct modeling, that the account of this component would uniformly reduce the required electron flux $F_{MeV}$ by a factor of two in the entire energy range without noticeable effect on other fitting parameters and primarily causing no changes in spectral indices.

In Extended Data Figure 2 we plot as points with error bars the normalized time histories of the $F_0 > 300\,\mathrm{keV}$ fluxes in the power-law extension of the electrons that produce the hard X-rays (red points) and that of the MeV-peaked electron population $F_{MeV}$ (blue points). Both populations rise at about the same time, but the MeV electron flux peaks about 1 minute earlier and falls more rapidly than does the power-law extension flux. The Figure also shows the microwave light curves integrated over ROIs 1 and 2 (red curve) and ROI 3 (blue curve). The blue curve displays a remarkable correspondence to the MeV-peaked $\gamma$-ray component, especially during the decay phase, while the red curve is better correlated with the normal power-law component, suggesting that these two $\gamma$-ray components arise in spatially distinct regions of the flare. This strongly suggests a causal relationship between the microwave and $\gamma$-ray emissions from the two populations. Note that the $\gamma$-ray to microwave light curve agreement is noticeably poorer in the rise phase than in the decay phase. A possible reason for that is rapidly changing flare parameters in the highly dynamic rise phase affecting the microwave emission, including the thermal number density due to chromospheric evaporation and decay of the magnetic field [2]. In the decay phase, such dynamics is less pronounced and the two emissions are better correlated.

## Spatially Resolved Microwave Data

The EOVSA spatially resolved microwave data[2] for a four-minute time range from 15:57 to 16:01 UT were described elsewhere [2, 10]. Here we extend the analyzed time range of this long-duration flare [24] to an 18 minute period from 15:48 to 16:06 UT covering the time period when the MeV-peaked component was observed in the $\gamma$-ray data; see Extended Data Fig. 2.

The model microwave spectra in Fig. 2 of the main text, shown in blue, are obtained for the following parameters: $n_{th} = 2.46 \times 10^{11}\,\mathrm{cm}^{-3}$, $n_{nth} \approx 5.4 \times 10^5\,\mathrm{cm}^{-3}$, $\delta = 2$, $B = 750\,\mathrm{G}$, $E_{\max} = 6\,\mathrm{MeV}$, and $E_{\min} = 2, 2.6, 3.6\,\mathrm{MeV}$, consistent with the results of the spectral fitting of the $\gamma$-ray emission, see above, and MCMC tests, see below. The one shown in red, representing ROI 2, is obtained for $n_{th} = 0.76 \times 10^{11}\,\mathrm{cm}^{-3}$, $n_{nth} \approx 3.5 \times 10^7\,\mathrm{cm}^{-3}$, $\delta = 2.6$, $B = 420\,\mathrm{G}$, $E_{\max} = 5\,\mathrm{MeV}$, and $E_{\min} = 20\,\mathrm{keV}$.

## MCMC Analysis of the rising microwave spectra

We investigated the statistical significance of the need for the MeV-peaked electron component to interpret the microwave emission from ROI 3 using Monte Carlo Markov

[2] The instrumental beam is $113''.7/f[\mathrm{GHz}] \times 53''.0/f[\mathrm{GHz}]$. A circular restoring beam of FWHM $87''.9/f[\mathrm{GHz}]$ was used, which is about $9''$ for 9.92 GHz.

Chain (MCMC) simulations. We used an open-source Python package *emcee* [25], to derive statistical distributions of, and cross-correlations between, the model fit parameters. This approach explores the full multidimensional space of the model fit parameters to both provide parameter distributions and reveal correlations between them. Because such analysis is time-consuming, we restrict our MCMC analysis to a subset of pixels in a single time frame, which inscribes ROI 3 and overlaps with ROI 1 and ROI 2. We have already performed a similar analysis elsewhere [10] but in that test we restricted $E_{\rm min} < 50$ keV in the nonthermal electron spectrum, which is a reasonable expectation for the flare-accelerated electron population [26] with a falling energy spectrum. Here, we remove this restriction to explicitly enable the code to find the MeV-peaked cases if favored by the data. Then, we restricted the microwave spectral range used for the MCMC analysis to high frequencies, $f > 8$ GHz, to focus on the high-frequency steep spectra. We note that having only a rising portion of the microwave spectrum (i.e., without the spectral peak) in a limited spectral range restricts the ability of the fit to uniquely recover all parameters, but the combination of the steep spectral slope and high brightness temperature turns out to be highly constraining regarding the shape of the involved nonthermal particle spectrum and the ambient plasma density.

Extended Data Figure 3 displays the MCMC analysis for an image pixel that belongs to ROI 1 (shown by a blue triangle in Fig. 1), which is not particularly steep. The tests favor a modest $E_{\rm min} \approx 70$ keV, which is consistent with reported [10] values. Other parameters are not well constrained; however, $E_{\rm min}$ shows a rather narrow distribution and thus is well constrained.

Pixels from ROI 3 display remarkably different behavior illustrated here by two representative pixels (blue cross and plus symbols in Fig. 1). The fits in Extended Data Figure 4 explicitly exclude $E_{\rm min}$ values below roughly 400 keV and favor $E_{\rm min} \approx$ 2.6 MeV. Extended Data Figure 5 shows a broader distribution of $E_{\rm min}$ permitting small values, but still favors a high value of $E_{\rm min}$ of about 1 MeV. In addition, the third panels from the right in the bottom row show clear correlation between $E_{\rm min}$ and the electron energy spectral index $\delta$. According to this correlation, low values of $E_{\rm min}$ are only permitted for proportionally smaller $\delta$, which emphasizes the dominant role of the MeV particles.

These MCMC simulations further help to firmly constrain another key parameter, the thermal number density. For pixels within ROI 3, this parameter displays a rather well constrained narrow distribution peaked at $n_{th} \approx (2-3) \times 10^{11}$ cm$^{-3}$. The physical ground of the gyro-emission sensitivity to the ambient density, already mentioned in the main text, is the Razin suppression [12] (also called the effect of density) of the emissivity and opacity in a dense plasma. In the considered spectral range, for the coronal magnetic field of about several hundred Gauss, the Razin suppression is efficient for $n_{th} \gtrsim 10^{11}$ cm$^{-3}$, which represents the lower bound of the ambient number density. The strength of inferring the thermal number density from the microwave data is that it is free from the line-of-sight integration effect unlike most other available diagnostics. The thermal number density affects the microwave spectrum by a modification of the refractive index of the waves in the source plasma, which does not explicitly depend on the source size or volume. Also, it does not depend on the plasma temperature unlike SXR or EUV diagnostics, but on the total electron number density only. The upper bound is $7.9 \times 10^{11}$ cm$^{-3}$, which would result in the plasma frequency of 8 GHz, because for larger plasma frequencies (larger densities) the emission is evanescent and cannot escape at the observed frequencies.

Other parameters reported by the MCMC simulations are not so well constrained. In the cases where the nonthermal number density is well-constrained, like in Extended Data Figure 5, the preferred values are around $n_{nth} \approx (2-4) \times 10^5$ cm$^{-3}$. The spectral index shows broad distribution from 2 to 8, i.e., poorly constrained, but fully consistent with the range of values reported in [9]. Therefore, the MCMC spectral model fitting of the microwave data within ROI 3 yields the source parameters quantitatively consistent with those obtained from the spectral fitting of the Fermi $\gamma$-ray data that assumed the thin-target bremsstrahlung emission model; see Methods above and [9]. In contrast, the microwave data do not favor a flat particle distribution extending up to 10-20 MeV implied by the inverse Compton interpretation of the $\gamma$-ray continuum, see fig. 8 in [9].

This analysis confirms the statistical significance of the conclusion that the microwave emission in ROI 3 is produced by the same MeV-peaked component as the one responsible for the corresponding $\gamma$-ray emission.

## Direct acceleration to MeV energies

The only fundamental microscopic force capable of accelerating charged particles from thermal to (much) higher energies is the electric force. Depending of the microscopic properties of the electric field, the acceleration mechanisms typically divide into two large groups—stochastic or direct/regular. Within a stochastic mechanism, a particle undergoes a sequence of accelerating and decelerating episodes resulting in a net acceleration. Typically, the energy gain in a single episode is small; thus, the higher the energy, the smaller the number of electrons capable of reaching this energy, resulting in a continuously falling spectrum with energy. The only possible exception to this rule is when the energy gain in a single accelerating episode is already large, which is effectively equivalent to an episode of direct acceleration. By this argument, to create an MeV-peaked electron population due to an acceleration mechanism, this acceleration must be direct.

Here we consider a theoretical model potentially capable of accelerating a small portion of available thermal electrons to MeV energies without a noticeable number of keV and sub-MeV electrons. Thermal electrons can only be accelerated to high energies by an ambient electric field $E$ and the portion of the electrons to be accelerated depends on the electric field strength; see the seminal paper [27] as a comprehensive reference. The steady-state in the plasma in the ambient electric field is determined by a balance between the accelerating electric force and decelerating drag force due to collisions between the plasma particles. The latter cannot exceed a certain maximum value, while the former can theoretically be arbitrarily strong resulting in unbalanced acceleration. Therefore, there is a critical electric field, called the Dreicer field $E_D$ [19], which demarcates regimes of highly efficient and less efficient acceleration. In a super-Dreicer field, $E \gtrsim E_D$, literally all available electrons "run away" and gain high energy because the collisional drag force becomes inefficient to slow them down in such a strong electric field. For a sub-Dreicer field, $E < E_D$, only rather fast electrons with velocities $v > v_{run} \approx v_T \left(E_D/E\right)^{1/2}$, where $v_T$ is the thermal velocity, are accelerated. In the Maxwellian distribution, $f(v) \propto \exp\left(-v^2/2v_T^2\right)$, only the corresponding portion $\eta_{run} = n_{run}/n_{th} \approx \exp\left(-v_{run}^2/2v_T^2\right)$ of the runaway electrons can be accelerated [27, 28] from the "tail" of the Maxwellian distribution; the smaller the electric field the smaller the fraction of accelerated electrons. From this consideration, it is apparent that to accelerate a small portion of electrons, such as implied by the analyzed here microwave and $\gamma$-ray data, we need either a rather small sub-Dreicer electric field or a very small filling factor of region(s) with super-Dreicer field.

Specifically, in the former case, to have $\eta_{run} \sim 10^{-6} \approx \exp(-14)$ as implied by the data, we need $v_{run}^2/2v_T^2 = E_D/2E = 14$ and, accordingly, $E = E_D/28$. For the source number density $n_{th} = 2.5 \times 10^{11}\,\mathrm{cm}^{-3}$ and a modest flare temperature $T = 10\,\mathrm{MK}$, the Dreicer electric field [12, 27, 28] is $E_D \approx 1.5 \times 10^{-3}\,\mathrm{V\,cm}^{-1}$; thus, the required (mean) electric field is $\langle E\rangle \approx E_D/28 \approx 5 \times 10^{-5}\,\mathrm{V\,cm}^{-1}$. To create an MeV electron population, this electric field must perform work on a majority of these runaway electrons equal to several MeV: $W = e\int \mathbf{E}\cdot d\mathbf{S} = e\,\langle E\rangle\, S \approx 2 \times 10^6\,\mathrm{eV}$, where $S$ is the length of the electron path. Given our estimated mean electric field, the required acceleration path is $S \sim 4 \times 10^{10}\,\mathrm{cm}$. This path length is to be compared with the source (ROI 3) size that is about $10^9$ cm—more than one order of magnitude shorter than the required path length. The only known physical process capable of holding a particle trajectory much longer than the source size is spatial diffusion, i.e., a random walk of the particle. This random walk is inconsistent with the requirement of the considered model that the electron motion is regular rather than stochastic. Therefore, the direct acceleration distributed with a filling factor about one over the source volume seemingly cannot account for the MeV-peaked electron population.

The latter case of the super-Dreicer electric field implies the volume filling factor of the accelerating region(s) of the order of $10^{-6}$, which implies the source size $L < 10^7$ cm. In such a case, all electrons will be accelerated in this small volume, giving a volume-averaged acceleration efficiency $\eta \sim 10^{-6}$. To gain the energy of $W = eEL \approx 2 \times 10^6\,\mathrm{eV}$ would require $E \gtrsim 0.1\,\mathrm{V\,cm}^{-1}$, roughly two orders of magnitude above the Dreicer field. This value is comparable with a roughly estimated inductive electric field for ROI 3, $\mathbf{E} \sim \mathbf{v}\times\mathbf{B} \sim 1\,\mathrm{V\,cm}^{-1}$, assuming that $v$ is about 0.1% of the Alfven speed. However, if this inductive electric field plays a role of the accelerating electric field, it remains unclear, why this is happening in only a tiny fraction $\sim 10^{-6}$ of ROI 3. Nevertheless, it is difficult to rule out this option based on the available data, although the data do not offer any signature of such a compact source. To be consistent with observations, this model must necessarily include very efficient electron transport to quickly spread the MeV electrons from the acceleration region with the size of $10^7$ cm over the source volume with the size of $10^9$ cm, while at the same time preventing their escape from the region. Thus; the electron transport must play a key role in shaping the observed source of the MeV-peaked electrons.

## Data availability:

All original EOVSA data are maintained in the EOVSA website at http://www.ovsa.njit.edu. Original EOVSA data used for this study are available at http://www.ovsa.njit.edu/fits/IDB/20170910/IDB20170910155625/. Fully processed EOVSA spectral imaging data in IDL save format can be downloaded from http://ovsa.njit.edu/publications/fleishman_ea_science_2019/data/ for the four-minute peak phase and https://ovsa.njit.edu/publications/fleishman_ea_natastro_2025/data/ for a longer, 18 minutes, duration. *Fermi*/GBM spectrum data file (.PHA) could be accessed at https://heasarc.gsfc.nasa.gov/FTP/fermi/data/gbm/daily/2017/09/10/current/ and Fermi/GBM response matrix for the event of interest could be accessed at http://hesperia.gsfc.nasa.gov/fermi/gbm/rsp/. Alternatively, data files could be retrieved through OSPEX.

## Code availability

All codes we use in this study are based on publicly available software packages: `GSFIT` is available in the community-contributed SolarSoftWare repository, under the packages category, at www.lmsal.com/solarsoft/ssw/packages/gsfit/; the open-source MCMC code is documented in [25], OSPEX spectral analysis software documentation is available at https://hesperia.gsfc.nasa.gov/ssw/packages/spex/doc/ospex_explanation.htm.

# Declarations

## Corresponding author

Correspondence to Gregory Fleishman gfleishm@njit.edu.

## Funding:

This work was supported in part by NSF grants AGS-2334931 and AGS-2425102 and NASA grants 80NSSC24K1242 and 80NSSC23K0090 to New Jersey Institute of Technology.

## Acknowledgments:

The authors thank Dr. G. Share for discussion of several aspects of this study and Prof. A. Sirenko for useful comments.

## Author contributions:

G.D.F performed modeling of the microwave emission for various particle distributions including the MeV-peaked component and wrote the draft manuscript; I.O analyzed Fermi $\gamma$-ray data including spectral model fitting and performed modeling of the $\gamma$-ray emission taking into account electron-electron bremsstrahlung; G.M.N performed spectral fitting of the rising microwave spectra and made comparative analysis of the steep and shallow spectra; B.C obtained self-calibration microwave data for the 18 minute time range employed in this study and applied the MCMC methodology to constrain the minimal energy in the electron spectra; S.Y performed time-domain correlation analysis of the microwave and $\gamma$-ray emissions; D.E.G led the construction and commissioning of the EOVSA, developed the observational strategy and calibration for microwave spectroscopy; and provided raw input data for the self-calibration. All authors discussed the interpretation of the data, contributed scientific results, and helped prepare the paper.

## Competing interests:

The authors declare that they have no competing financial interests.

# Extended Data Figures and Tables

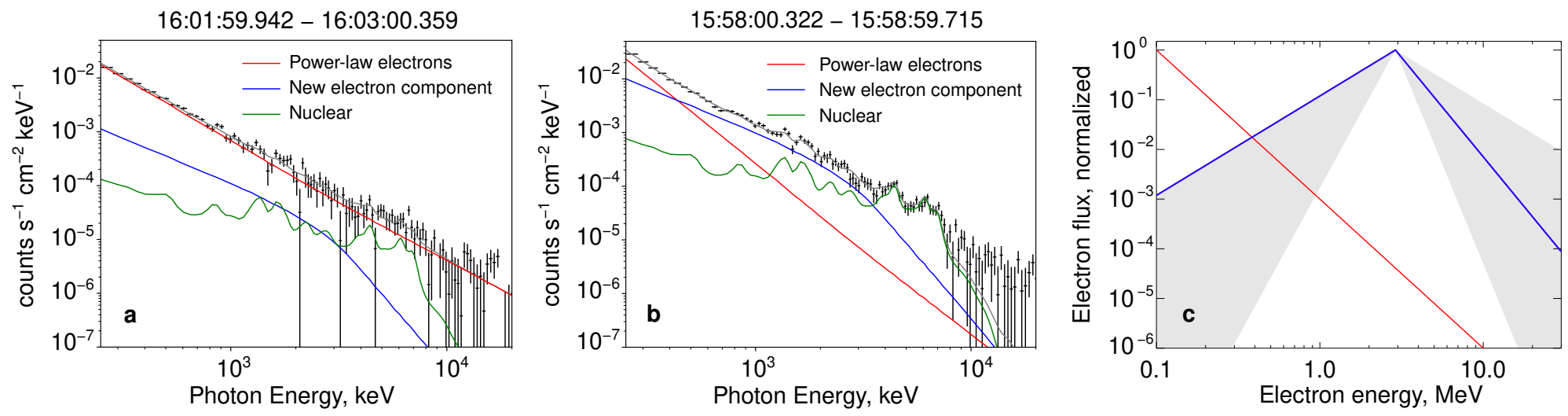

**Extended Data Fig. 1** ***Fermi*/GBM spectra of two one-minute accumulations during the 2017 September 10 flare and their spectral fits.** (a) Counts spectrum near the peak of the MeV electron emission. (b) Counts spectrum at a time when there is still strong power-law electron emission, but the MeV emission is weak. (c) The electron energy spectrum (blue lines) needed to generate thin-target bremsstrahlung is shown in blue in panels (a, b). The light gray polygons show uncertainties in the slopes of this spectrum. For reference, the red line shows the normalized spectrum of electrons responsible for the standard power-law X-ray / $\gamma$-ray component shown in red in panels (a, b).

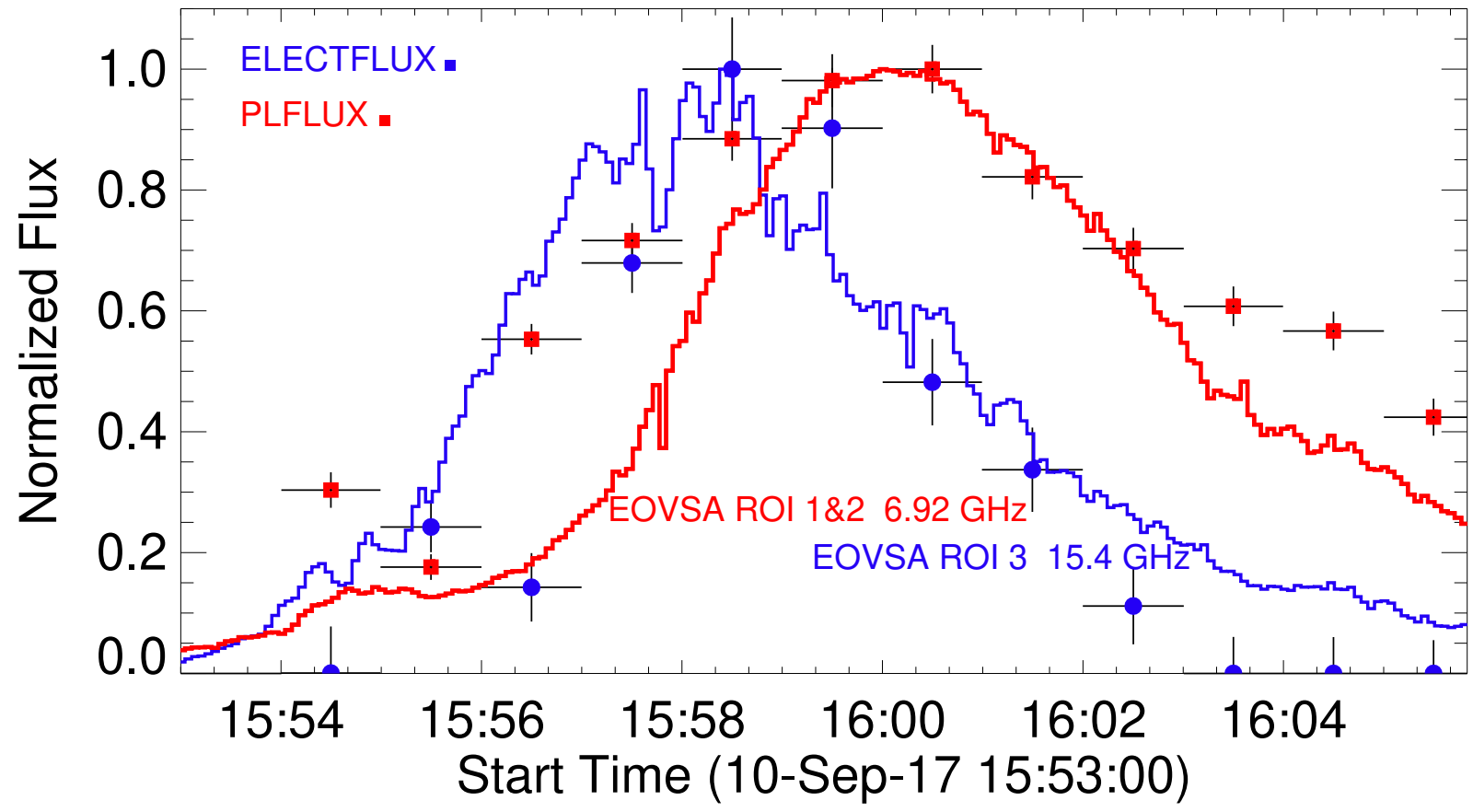

**Extended Data Fig. 2 Time histories of the power-law and MeV-peaked electron populations in the 2017 September 10 flare observed by the *Fermi*/GBM.** Red points show the photon flux above 300 keV of the standard power-law component of the flare-accelerated electrons, while the blue circles show the electron flux above 300 keV of the MeV component derived from the thin-target bremsstrahlung fit. The red line shows microwave emission at 7.92 GHz integrated over ROI 1 and ROI 2. The blue line shows microwave emission at 15.9 GHz integrated over ROI 3.

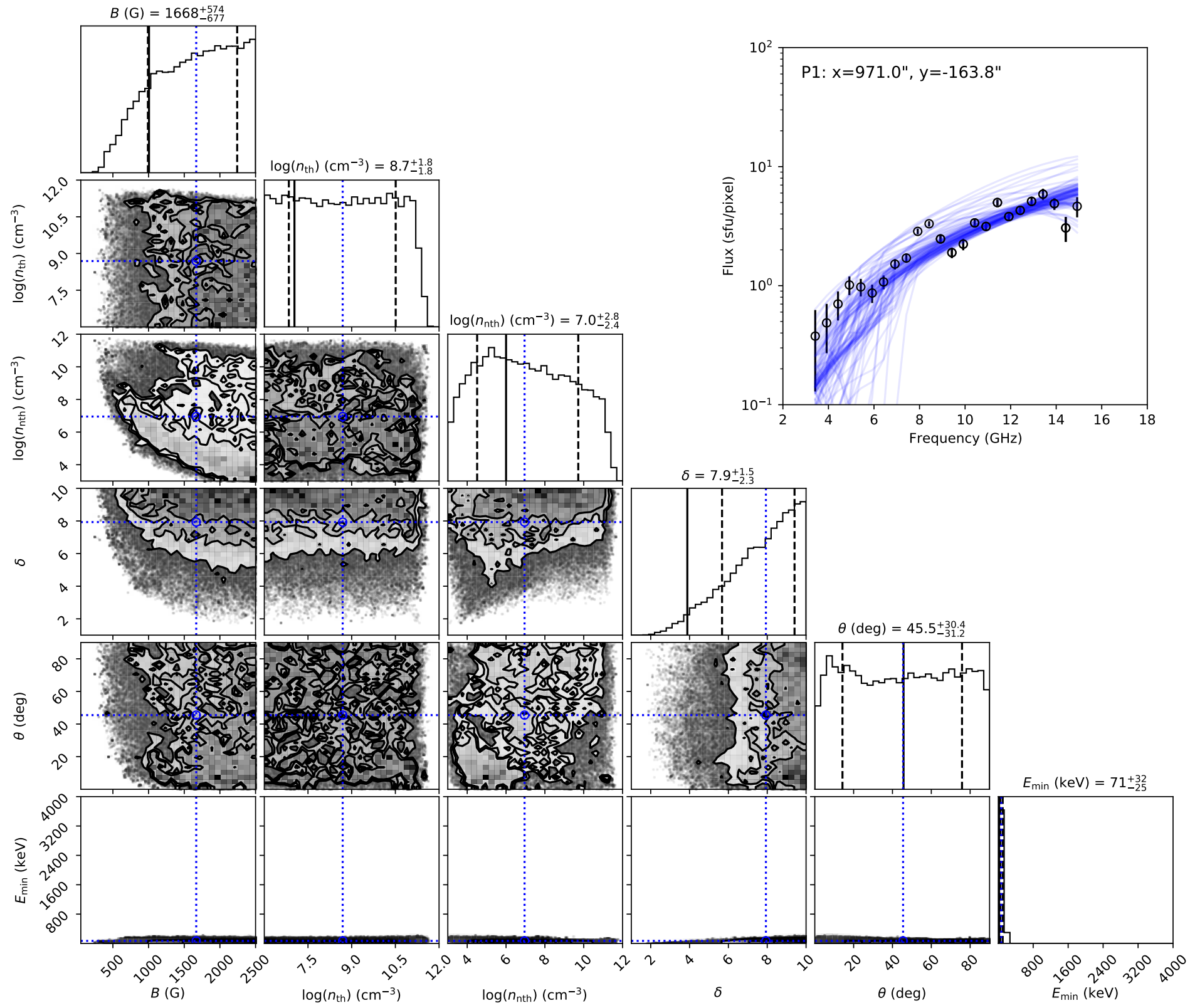

**Extended Data Fig. 3 MCMC probability distributions of the fit parameters for an example pixel in ROI 1**. The results are for pixel P1 located at $x = 971.0''$ and $y = -164''$ outlined by small blue triangle in Fig. 1. Solid black horizontal/vertical lines in each panel indicate the best-fit values from the GSFIT minimization. Dotted blue horizontal/vertical lines mark the median values of the MCMC probability distributions. Dashed lines in the histograms along the diagonal indicate $\pm$1-sigma standard deviation of a given parameter. Off-diagonal panels show correlations between all possible pairs of parameters shown as two-dimensional histograms of the probability distributions. The contour levels represent 39.3%, 60%, and 80% of the maximum. The outer contour level is selected to represent approximately the 1-sigma region of a 2D Gaussian distribution $(1 - e^{-0.5})$.

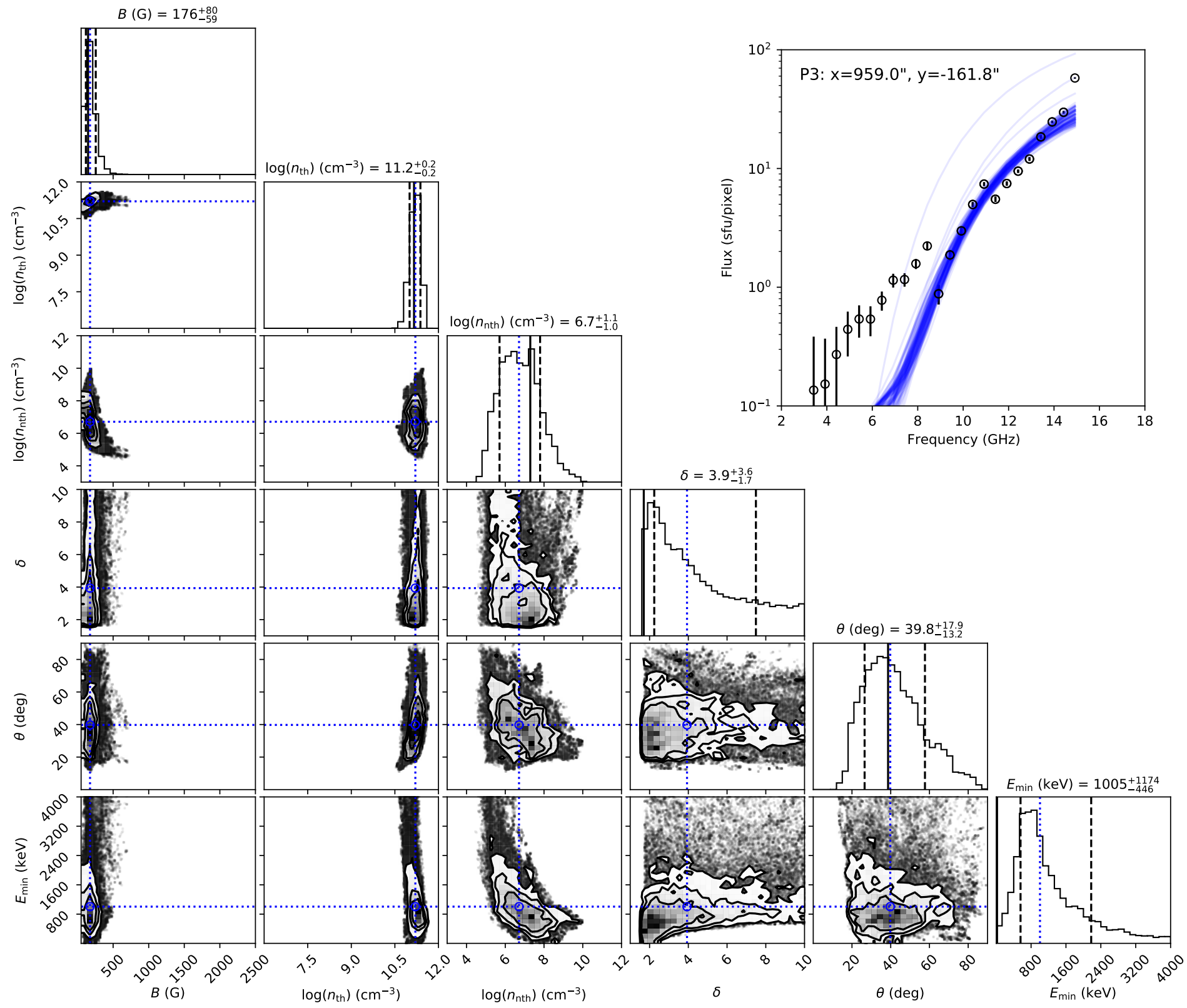

**Extended Data Fig. 4 MCMC probability distributions of the fit parameters for another pixel in ROI 3**. The figure layout is identical to Extended Data Fig. 3, but showing the parameters for pixel P3 located at $x = 959.0''$ and $y = -162''$ as marked by the small blue plus sign in Fig. 1.

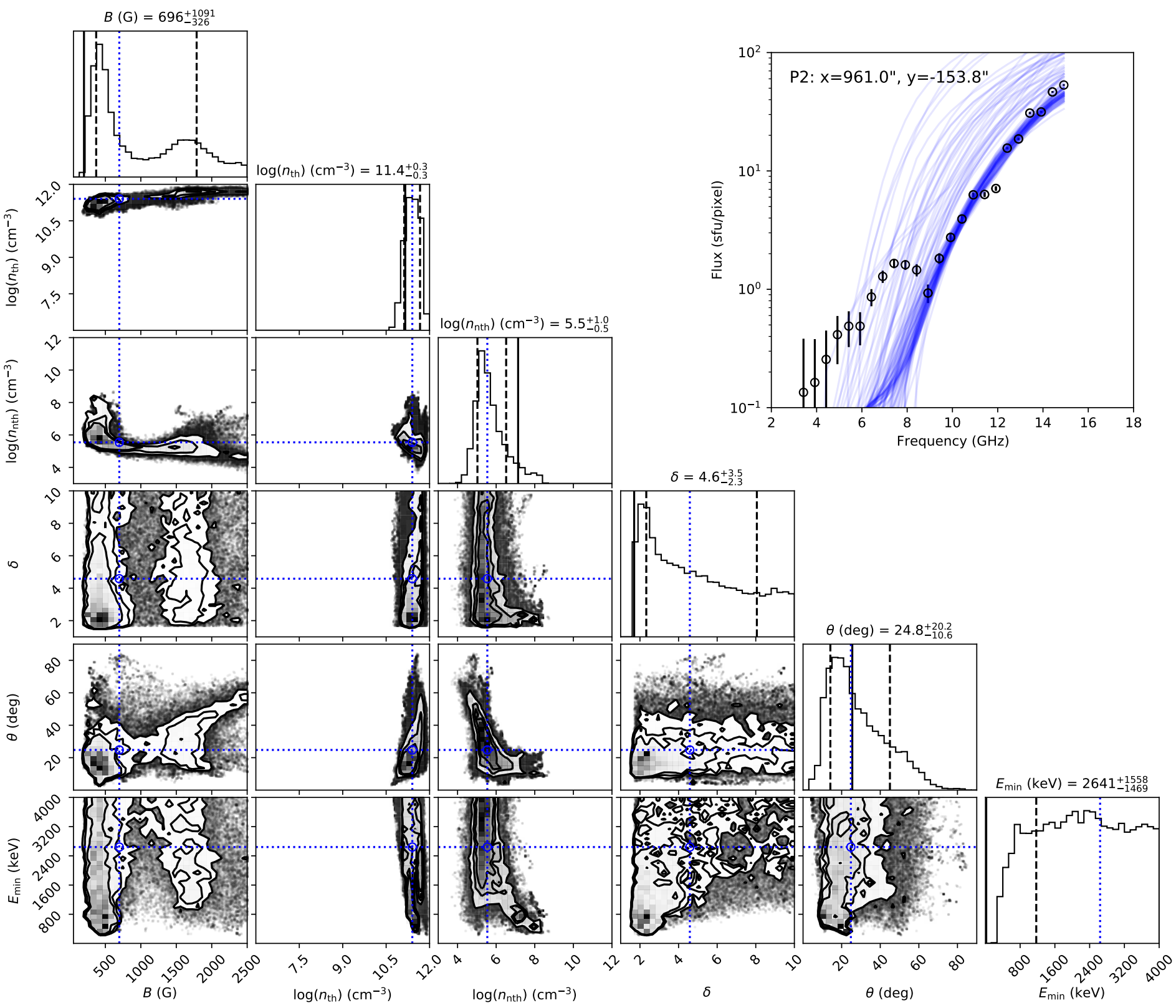

**Extended Data Fig. 5 MCMC probability distributions of the fit parameters for an example pixel in ROI 3**. The figure layout is identical to Extended Data Fig. 3, but showing the parameters for pixel P2 located at $x = 961.0''$ and $y = -154''$ as marked by the small blue cross in Fig. 1.

## Supplementary Video

**Supplementary Video S1. Evolving map of the microwave spectral index $\beta$.** This Supplementary Video demonstrates evolution of the spectral index $\beta$ of the spatially resolved microwave flux $F_f \propto f^\beta$ at the locations, where the microwave brightness temperature is above $10^9$ K at least at one of the observing frequencies. The video demonstrates that the steep spectra are concentrated at/around ROI 3 and they appear and then disappear during the considered time episode.